\documentclass[structabstract]{aa}  
\usepackage{graphicx}
\usepackage{natbib}
\usepackage{lscape}
\usepackage{multirow}
\usepackage{txfonts}
\begin{document}
\def\gsim{\;\lower.6ex\hbox{$\sim$}\kern-6.7pt\raise.4ex\hbox{$>$}\;}
\def\lsim{\;\lower.6ex\hbox{$\sim$}\kern-6.7pt\raise.4ex\hbox{$<$}\;}

   \title{Anomalous Cepheids in the Large Magellanic Cloud: \thanks
     {Based on observations released by the Optical Gravitational
       Lensing Experiment (OGLE-III).}}

   \subtitle{Insight into their origin and connection with the star formation history.}

\author{G. Fiorentino\inst{1}
\&
 M. Monelli\inst{2,3}}

\institute{INAF- Osservatorio Astronomico di Bologna, via Ranzani 1, 40127, Bologna, Italy.\\
\email{giuliana.fiorentino@oabo.inaf.it}
\and
Instituto de Astrof\'{i}sica de Canarias, Calle Via Lactea s/n, E38205 La Laguna, Tenerife, Spain.
\and
Departmento de Astrof\'{i}sica, Universidad de La Laguna, E38200 La Laguna, Tenerife, Spain
}

   \date{Received Nov 09, 2011; accepted XX XX, 2011}

 
  \abstract
{
The properties of variable stars can give independent constraints on the star 
formation history of the host galaxy, by determining the age and
metallicity of the parent population.
}
{
We investigate the pulsation properties of 84 Anomalous Cepheids (ACs) detected by the 
OGLE-III survey in the
Large Magellanic Cloud (LMC), in order to understand the formation mechanism and the 
characteristics of the parent population they came from.
}
{
We used an updated theoretical pulsation scenario to derive the mass and the pulsation mode 
of each AC in the sample. We also used a Kolmogorov-Smirnov test to analyze 
the spatial distribution of the ACs, in comparison with that of other groups of 
variable stars, and connect their properties with the star formation history of the LMC. 
}
{
We find that the mean mass of ACs is $1.2 \pm 0.2 M_{\odot}$. We show that ACs do not 
follow the same spatial distribution of classical Cepheids. This and the difference in
their period-luminosity relations further support the hypothesis
that ACs are not the 
extension to low luminosity of classical Cepheids. The spatial distribution of ACs is 
also different from that of bona-fide tracers of the old population, such as RR Lyrae 
stars and population II Cepheids. We therefore suggest that the majority of ACs in 
the LMC are made of intermediate-age (1-6\,Gyr), metal-poor single stars.
Finally, we investigate the relation between the frequency of ACs and the luminosity
of the host galaxy, disclosing that purely old systems follow a very tight relation
and that galaxies with strong intermediate-age and young star formation
tend to have an excess of ACs, in agreement with their hosting
ACs formed via both single and binary star channels.} 
{}

\authorrunning{G.~Fiorentino \& M.~Monelli}
\titlerunning{ACs in LMC}

   \maketitle
%

	\section{Introduction}\label{sec:intro}

Anomalous Cepheids (ACs) are short-period variable stars (from a few hours to two days), 
located on the colour-magnitude diagram $\sim 1$ mag above the horizontal branch, at 
colours similar to those of RR Lyrae stars \citep{bono97c}. Their properties can be explained 
by assuming that they are relatively massive ($1-2 M_{\odot}$), core-He burning stars in a very 
low-metallicity regime that ignite helium under partial electron-degeneracy 
conditions \citep[$Z \lsim 0.0008$,][]{renzini77,castellani95,bono97c,caputo04,fiorentino06}.  

ACs have been observed in many nearby Local Group dwarf galaxies that were surveyed 
for short time variability, independently of the morphological type. Typically, nearby dwarfs
host a few of them. They have been observed in {\it purely old systems}
($age \gsim 9-10$\,Gyr), such as in the dwarf spheroidal (dSph) satellites
Sculptor \citep[3 ACs,][]{smith86, kaluzny95}, Sextans
\citep[6,][]{mateo95}, LeoII \citep[4,][]{siegel00}, Ursa Minor 
\citep[7,][]{nemec88}, Draco \citep[10,][]{harris98,kinemuchi08},
and also in isolated dSphs, like Cetus (8, \citealt{bernard09}, \citealt{monelli12a}),
and Tucana \citep[6,][]{bernard09}. However, they have also been found both in 
 dSph galaxies with a large  {\it intermediate-age population}
(hereinafter, $1-6$\,Gyr), such as Fornax \citep[17,][]{bersier02}, Carina \citep[15,][]{dallora03},
and LeoI \citep[15,][]{hodge78} and in gas-rich dwarfs such as Phoenix
\citep[12,][]{gallart04}. The cases of Leo~A and 
NGC~6822 are not as clear, for the coexistence of ACs and short-period classical Cepheids
\citep{hoessel94, dolphin02, baldacci05}. 
Furthermore, some have been identified in four satellites of M31
\citep{pritzl02, pritzl04, pritzl05} for which the star formation
history and the the population content is still uncertain.
Recently, three ACs have been proposed as candidates in the ultrafaint dwarf CVenI \citep{kuehn08},
However, nearby galaxies, especially the spheroidal satellites of the Milky
Way, typically occupy large areas of the sky, so that probably the actual census is 
far from complete. On the other hand, ACs are very rare in globular clusters: 
so far, only one candidate has been confirmed in the metal-poor ($\mathrm{[Fe/H]} \sim -2$~dex)
NGC~5466 \citep[][]{zinn76}, and a few
others have been suggested \citep{corwin99, arellano08, kuehn11}.

The largest sample of ACs in nearby systems (83) has been collected for the LMC in the framework
of the OGLE-III project \citep[Optical Gravitational Lensing Experiment,][]{udalski08}. In a series
of papers devoted to the analysis of thousands of variable stars in the LMC (classical Cepheids,
\citealt{soszynski08b}; ACs and population II Cepheids, \citealt{soszynski08c}; RR Lyrae, \citealt{soszynski09a}; 
long-period 
variables, \citealt{soszynski09b}), the OGLE-III experiment released the $V$, $I$ photometry, 
pulsation properties, and well-sampled light curves for variables with periods
from a few hours to hundreds of days. In this paper, we take advantage of this rich data base
to investigate the nature of ACs within the global picture depicted by the star 
formation history and chemical evolution available nowadays 
\citep[e.g. ][]{carrera08a,harris09,saha10,cioni11, carrera11} for the LMC.

In fact, even though ACs identify a population of stars in a well-defined mass range, the 
origin of their progenitors is still being debated. If they are the result of the evolution
of single stars, then they trace a relatively young event in the star formation, which occurred
from $\sim ~1$ to $6$\,Gyr ago \citep{demarque75,norris75,castellani95,caputo99}. On the other hand, 
it has been
proposed that they are the evolution of blue straggler stars (BSS), formed via mass-exchange in 
binary systems \citep{renzini77,sills09} that survived in low-density
environments, so ACs are
 tracers of the old, metal-poor population. The discovery of 83 ACs in a relatively metal-rich 
 ($Z=0.008$) environment such as that of the LMC makes this picture even more puzzling.

In this work, we present the approach based on pulsation models built {\it ad hoc} 
for this class of variable stars \citep{marconi04,caputo04,fiorentino06} to constrain
the pulsation mass of individual stars. This, in turn,
can be related to the star formation history (SFH) of the host system, as already successfully done to
all the ACs discovered so far in nearby galaxies \citep[][]{caputo04,monelli12a,fiorentino12a}.

This paper is organized as follows. In Sect.~\ref{sec:data} we give a full
description of the sample selected from the OGLE-III survey of the LMC and
of the Fourier analysis needed to measure the pulsation amplitude in
V-band. In Sect.~\ref{sec:comparison} we compare the OGLE-III data
with the theoretical boundaries of the instability strip
\citep{marconi04}. In Sect.~\ref{sec:masses}, we describe
the theoretical method used to simultaneously constrain the pulsation modes and the
masses. Section~\ref{sec:spatial} presents the radial distribution of different samples of
variable stars in the LMC. The discussion and the conclusions close the paper. 


	\section{The OGLE-III sample}\label{sec:data}

\begin{figure}
\includegraphics[width=8cm]{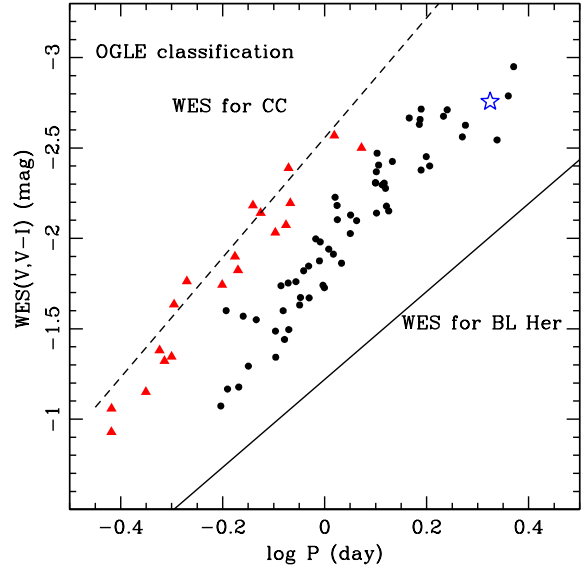}
\caption{Wesenheit ($V, V-I$) plane. Dots and triangles represent
  fundamental and first-overtone anomalous Cepheids as defined by
  OGLE-III \citep[see][]{soszynski08c}. The star represents one P2C
  candidate from   OGLE-III that we have included in our sample. The
  solid line represents the theoretical prediction for BL Her type 
  stars from \citet{dicriscienzo07}, while the dashed line is the 
  observed Wesenheit relation for fundamental-mode classical Cepheids 
  \citep{soszynski08a}.
\label{fig:wes}}
\end{figure}

We selected the 83 ACs identified in the short period Cepheids catalogue presented in
\citet{soszynski08c}. The classification of the variable type is mostly
based on the period-luminosity relation (PL), as detailed in \citet{soszynski08a,
soszynski08c}. In particular, the Wesenheit\footnote{The Wesenheit index is defined 
as $W = V - 2.54\times(V-I)$, where $V$ and $I$ are the apparent magnitudes.} index 
is a good diagnostic, because it includes the colour information (and therefore the
temperature) of the star without being affected by the reddening evaluation.
Using this plane, it is straightforward for separating Cepheids (classical and anomalous)
from population II Cepheids (hereinafter P2C, which include BL Her, W Vir, and RV Tau), 
because the latter have, at a given period, significantly fainter magnitudes 
\citep[][and references therein]{fiorentino06}. From the Wesenheit plane, it clearly 
emerges that and classical Cepheids occupy different regions. In 
particular, ACs are located at an intermediate position between classical Cepheids and 
P2C. However, the unambiguous distinction between anomalous and classical Cepheids is 
hampered by the fact that the first-overtone (FO) AC fall on the extension of the 
period-luminosity relation to shorter periods of the fundamental (F) classical Cepheids
\citep[$period \lsim 2.5$~d][]{caputo04,soszynski08c}. For these stars, the shape of 
the light curve is the most efficient diagnostic to separate the two groups. This 
picture is even more complicated by certain degeneracy existing with the RR Lyrae 
stars as well. Therefore, while ACs with periods longer than 1~d can be safely identified, 
in the short-period regime ($<1$~d), brighter RR Lyrae stars and ACs can overlap on
the period-magnitude plane. Moreover, the intrinsic similarities between the light 
curves of these two groups make the distinction uncertain.

The final classification from OGLE-III includes 83 ACs. On the basis of their 
location in the Wesenheit plane, 62 were classified as F and 21 as FO pulsators. 
We discuss the importance of the mode classification more fully in Sect.~\ref{sec:masses}.
To this sample, we added one star classified by OGLE-III as P2C (namely V-166), because it shares the pulsation properties of ACs, as we discuss in 
the next section. Figure~\ref{fig:wes} shows the ($V$, $V-I$) Wesenheit relation for 
the 84 stars. For comparison, we also show the theoretical relation for BL Her stars given by 
\citet{dicriscienzo07}, which are the subsample of the P2C class with 
a short period ($P \le 4$~d). The Wesenheit relation for fundamental-mode classical 
Cepheids has been taken from \citet{soszynski08a}.

In Table~\ref{tab:tab1} we summarize some of the main properties provided by the
OGLE-III catalogue: the first eight columns give, for each variable star, the identification
number,
the position (right ascension and declination), the mean $V$ and $I$ magnitudes, 
the period, the amplitude in the $I$ band ($A_I$), and the mode classification.
The $V$-band amplitude, $A_V$, was not given because the number of phase points in the 
V-band (27 to 138) is typically a factor $\sim 10$ smaller than in $I$ (319 to 1299).
Nevertheless, the minimum number of data points in the $V$ band is 27 epochs,
ensuring well-sampled light curves. For this reason, we performed the Fourier 
analysis on V-band photometry to give an independent estimation of the pulsation
parameter and provide the amplitude $A_V$. Taking advantage of the accurate periods 
given from OGLE-III catalogue, we estimated $A_V$ using the GRATIS code \citep[developed by P. 
Montegriffo, see][and reference therein]{fiorentino10a}. The results are reported in 
columns 9 and 10 of Table~\ref{tab:tab1}. The comparison with the OGLE-III mean magnitude 
reveals that this difference $\Delta <V>$ is always less than about 0.05 mag. The total error 
budget on our evaluation gives $\sigma (A_V) \lsim 0.1$~mag, thus perfectly suitable
for the goals of the present analysis. The star V-82 was excluded because no 
V-band photometry is available.

\begin{table*}
\caption{The anomalous Cepheids from the OGLE-III sample, including 1 star 
 (last row) classified by OGLE-III as P2C (see text for details.)
\label{tab:tab1}}
\centering
\scriptsize
\begin{tabular}{lccccccc|cccccl}
\hline
\multicolumn{8}{c|}{OGLE-III} & \multicolumn{6}{c}{This work} \\
\hline
\hline
ID  & R.A.          & Decl.         & V      & I   & P & $A_I$ & $mode_{OGLE}$ & $<V>$ & $A_V$ & $E(V-I)$ & $M_{F}$ & $M_{FO}$ &  $mode_{mass}$\\
    & {\itshape J2000} & {\itshape J2000} & {\it mag} & {\it mag}  & {\it d} & {\it mag}   &      & {\it mag} & {\it mag} & {\it mag} & $M_{\odot}$ & $M_{\odot}$ &  \\
\hline
\hline
001 & 04:37:18.96 & -69:49:11.5 & 18.676 & 18.018 & 0.8502326 &  0.652 &   F  & 18.692 &  0.843 &  0.110 &  0.9  &  $-$  &   F   \\
002 & 04:41:56.22 & -66:51:59.1 & 18.268 & 17.621 & 0.9766401 &  0.511 &   F  & 18.264 &  0.838 &  0.060 &  1.1  &  $-$  &   F   \\
003 & 04:43:16.03 & -69:07:57.6 & 18.794 & 18.313 & 0.3817797 &  0.529 &  FO  & 18.787 &  0.880 &  0.110 & (1.8) &  1.0  &   F** \\
004 & 04:47:21.63 & -70:22:12.0 & 17.719 & 17.018 & 1.8618275 &  0.637 &   F  & 17.739 &  0.959 &  0.140 &  1.0  &  $-$  &   F   \\
005 & 04:51:11.62 & -69:00:33.0 & 18.526 & 17.858 & 0.9321588 &  0.243 &   F  & 18.527 &  0.382 &  0.160 &  1.2  &  $-$  &   F   \\
006 & 04:52:34.51 & -69:43:27.5 & 17.671 & 17.057 & 0.8495758 &  0.389 &  FO  & 17.671 &  0.682 &  0.140 &  $-$  &  1.7  &  FO   \\
007 & 04:57:31.48 & -70:15:53.6 & 18.247 & 17.688 & 0.8963988 &  0.703 &   F  & 18.143 &  1.180 &  0.080 &  1.1  &  $-$  &   F   \\
008 & 04:58:24.59 & -71:05:11.5 & 17.911 & 17.300 & 0.7490681 &  0.262 &  FO  & 17.912 &  0.425 &  0.100 &  $-$  &  1.6  &  FO   \\
009 & 04:58:51.29 & -67:44:23.9 & 17.913 & 17.344 & 0.8000803 &  0.366 &  FO  & 17.896 &  0.597 &  0.060 & (2.2) &  1.2  &   F** \\
010 & 04:59:00.10 & -68:14:01.1 & 18.723 & 18.068 & 0.8342013 &  0.705 &   F  & 18.628 &  1.082 &  0.110 &  0.8  &  $-$  &   F   \\
011 & 04:59:38.09 & -70:37:45.5 & 18.254 & 17.671 & 0.9985904 &  0.561 &   F  & 18.299 &  0.843 &  0.035 &  1.0  &  $-$  &   F   \\
012 & 04:59:49.43 & -69:33:12.4 & 18.663 & 17.969 & 0.8289912 &  0.445 &   F  & 18.679 &  0.719 &  0.100 &  1.1  &  $-$  &   F   \\
013 & 04:59:56.62 & -70:42:24.9 & 18.488 & 17.963 & 0.5009227 &  0.446 &  FO  & 18.480 &  0.737 &  0.030 & (1.8) &  1.1  &   F** \\
014 & 05:00:08.26 & -67:54:04.3 & 17.241 & 16.639 & 2.2913456 &  0.580 &   F  & 17.263 &  0.787 &  0.050 &  1.1  &  $-$  &   F   \\
015 & 05:00:59.32 & -69:33:24.9 & 17.587 & 16.962 & 1.1808787 &  0.383 &  FO  & 17.578 &  0.540 &  0.090 & (2.2) &  1.1  &   F** \\
016 & 05:01:36.69 & -67:51:33.1 & 17.448 & 16.926 & 1.5456697 &  0.763 &   F  & 17.450 &  1.197 &  0.055 &  1.2  &  $-$  &   F   \\
017 & 05:02:03.13 & -68:09:30.4 & 18.190 & 17.585 & 0.9299953 &  0.550 &   F  & 18.185 &  0.897 &  0.060 &  1.2  &  $-$  &   F   \\
018 & 05:02:36.36 & -69:40:12.2 & 18.213 & 17.562 & 1.0185719 &  0.476 &   F  & 18.229 &  0.798 &  0.085 &  1.2  &  $-$  &   F   \\
019 & 05:03:17.71 & -68:49:33.5 & 17.891 & 17.414 & 0.9094057 &  0.839 &   F  & 17.895 &  1.295 &  0.050 &  1.3  &  $-$  &   F   \\
020 & 05:05:54.54 & -70:43:30.3 & 18.359 & 17.998 & 0.3819013 &  0.491 &  FO  & 18.362 &  0.768 &  0.060 &  $-$  &  1.2  &  FO   \\
021 & 05:06:37.49 & -68:23:40.3 & 17.827 & 17.188 & 1.2958429 &  0.750 &   F  & 17.821 &  1.166 &  0.070 &  1.2  &  $-$  &   F   \\
022 & 05:06:45.08 & -70:52:44.7 & 18.378 & 17.796 & 0.6409626 &  0.484 &   F  & 18.400 &  0.754 &  0.070 &  1.7  &  $-$  &   F   \\
023 & 05:07:52.42 & -68:50:28.8 & 17.763 & 17.194 & 0.7234331 &  0.400 &  FO  & 17.761 &  0.650 &  0.080 &  $-$  &  1.8  &  FO   \\
024 & 05:08:44.05 & -68:46:01.2 & 17.986 & 17.679 & 0.7944643 &  0.471 &   F  & 18.000 &  0.747 &  0.110 &  $-$  &  $-$  &  n.c. \\
025 & 05:09:55.67 & -70:03:08.2 & 18.413 & 17.904 & 0.4744618 &  0.354 &  FO  & 18.417 &  0.574 &  0.065 & (2.3) &  1.3  &   F** \\
026 & 05:10:42.62 & -68:48:19.6 & 17.483 & 16.816 & 1.7387452 &  0.607 &   F  & 17.483 &  0.987 &  0.080 &  1.3  &  $-$  &   F   \\
027 & 05:10:45.77 & -67:10:05.6 & 17.558 & 16.956 & 1.2670644 &  0.973 &   F  & 17.550 &  1.786 &  0.070 &  1.4  & (1.0) &  FO*  \\
028 & 05:13:10.02 & -68:46:11.9 & 18.322 & 17.574 & 0.5992535 &  0.332 &  FO  & 18.320 &  0.672 &  0.060 &  $-$  &  $-$  &  n.c. \\
029 & 05:13:33.58 & -67:33:42.2 & 18.745 & 18.120 & 0.8017029 &  0.441 &   F  & 18.745 &  0.659 &  0.100 &  0.9  &  $-$  &   F   \\
030 & 05:15:07.61 & -68:04:21.4 & 17.685 & 17.258 & 0.6670589 &  0.538 &  FO  & 17.685 &  0.877 &  0.085 &  $-$  &  1.4  &  FO   \\
031 & 05:15:32.98 & -68:10:29.4 & 17.727 & 17.215 & 0.8397205 &  0.401 &  FO  & 17.728 &  0.689 &  0.115 &  $-$  &  1.1  &  FO   \\
032 & 05:15:56.14 & -69:01:29.1 & 17.780 & 17.167 & 1.3160220 &  0.425 &   F  & 17.778 &  0.651 &  0.050 &  1.3  &  $-$  &   F   \\
033 & 05:16:08.01 & -70:39:16.5 & 17.308 & 16.616 & 2.3470013 &  0.186 &   F  & 17.305 &  0.287 &  0.050 &  1.3  &  $-$  &   F   \\
034 & 05:17:11.15 & -69:58:33.1 & 18.474 & 17.874 & 0.7342975 &  0.312 &   F  & 18.477 &  0.561 &  0.120 &  1.4  &  $-$  &   F   \\
035 & 05:18:22.19 & -69:03:38.4 & 18.658 & 18.143 & 0.4460486 &  0.508 &  FO  & 18.658 &  0.793 &  0.080 & (1.8) &  1.1  &   F** \\
036 & 05:18:58.84 & -69:26:47.8 & 17.787 & 17.160 & 1.2579824 &  0.281 &   F  & 17.787 &  0.406 &  0.075 &  1.6  &  $-$  &   F   \\
037 & 05:19:16.66 & -70:11:58.4 & 17.743 & 17.132 & 1.2577403 &  0.794 &   F  & 17.703 &  1.099 &  0.080 &  1.4  &  $-$  &   F   \\
038 & 05:19:59.97 & -71:45:23.7 & 17.908 & 17.294 & 1.3352133 &  0.518 &   F  & 17.907 &  0.797 &  0.060 &  1.1  &  $-$  &   F   \\
039 & 05:20:44.46 & -69:47:46.5 & 18.245 & 17.660 & 0.9924071 &  0.621 &   F  & 18.249 &  0.908 &  0.050 &  1.0  &  $-$  &   F   \\
040 & 05:21:13.22 & -70:34:20.7 & 18.037 & 17.433 & 0.9605775 &  0.513 &   F  & 18.040 &  0.803 &  0.095 &  1.5  &  $-$  &   F   \\
041 & 05:21:14.35 & -70:29:39.5 & 18.200 & 17.625 & 0.8781418 &  0.548 &   F  & 18.203 &  0.854 &  0.095 &  1.3  &  $-$  &   F   \\
042 & 05:23:34.60 & -69:10:58.2 & 18.715 & 17.897 & 1.0790357 &  0.671 &   F  & 18.705 &  1.019 &  0.110 &  0.8  & (0.6) &  FO*  \\
043 & 05:24:35.40 & -68:48:22.6 & 18.593 & 17.913 & 0.5064704 &  0.356 &  FO  & 18.593 &  0.542 &  0.090 &  $-$  &  1.6  &  FO   \\
044 & 05:25:54.11 & -69:26:52.9 & 17.609 & 17.052 & 1.3085090 &  0.913 &   F  & 17.596 &  1.398 &  0.060 &  1.2  &  $-$  &   F   \\
045 & 05:26:24.68 & -68:57:55.0 & 18.976 & 18.325 & 0.6784311 &  0.453 &   F  & 18.977 &  0.804 &  0.110 &  0.9  &  $-$  &   F   \\
046 & 05:26:27.17 & -69:58:57.0 & 17.851 & 17.264 & 1.2637169 &  0.592 &   F  & 17.849 &  0.931 &  0.030 &  1.1  &  $-$  &   F   \\
047 & 05:27:05.27 & -71:23:33.4 & 17.482 & 16.881 & 2.1779846 &  0.493 &   F  & 17.475 &  0.769 &  0.090 &  0.9  &  $-$  &   F   \\
048 & 05:27:12.12 & -69:37:19.6 & 17.324 & 16.718 & 1.5458930 &  0.751 &   F  & 17.321 &  1.209 &  0.060 &  1.6  &  $-$  &   F   \\
049 & 05:28:03.57 & -69:39:15.2 & 18.451 & 18.011 & 0.6447960 &  0.451 &   F  & 18.455 &  0.682 &  0.050 &  1.2  &  $-$  &   F   \\
050 & 05:28:57.71 & -70:07:15.5 & 17.049 & 16.609 & 1.0446912 &  0.383 &  FO  & 17.053 &  0.591 &  0.060 &  $-$  &  1.5  &  FO   \\
051 & 05:30:14.24 & -68:42:31.1 & 18.845 & 18.200 & 0.7086059 &  0.318 &   F  & 18.842 &  0.559 &  0.100 &  1.0  &  $-$  &   F   \\
052 & 05:31:01.53 & -70:42:22.2 & 17.577 & 17.008 & 1.2625549 &  0.823 &   F  & 17.579 &  1.309 &  0.070 &  1.4  &  $-$  &   F   \\
053 & 05:31:06.20 & -68:43:45.3 & 17.299 & 16.738 & 1.8880987 &  0.721 &   F  & 17.289 &  1.128 &  0.110 &  1.3  &  $-$  &   F   \\
054 & 05:31:06.72 & -68:22:29.8 & 18.798 & 17.901 & 0.9802225 &  0.739 &   F  & 18.776 &  1.197 &  0.260 &  1.1  & (0.8) &  FO*  \\
055 & 05:31:41.11 & -68:44:37.7 & 17.603 & 17.011 & 1.6066649 &  0.850 &   F  & 17.549 &  1.266 &  0.080 &  1.1  &  $-$  &   F   \\
056 & 05:31:49.45 & -70:33:22.6 & 17.877 & 17.284 & 1.1240035 &  0.727 &   F  & 17.861 &  1.070 &  0.060 &  1.3  &  $-$  &   F   \\
057 & 05:31:49.88 & -70:46:30.0 & 17.455 & 16.813 & 1.7100082 &  0.686 &   F  & 17.437 &  1.307 &  0.070 &  1.2  &  $-$  &   F   \\
058 & 05:33:39.38 & -70:21:28.6 & 18.055 & 17.710 & 0.4852351 &  0.255 &  FO  & 18.056 &  0.321 &  0.080 &  $-$  &  1.1  &  FO   \\
059 & 05:34:14.81 & -68:23:43.0 & 20.004 & 18.724 & 0.8348234 &  0.717 &   F  & 20.011 &  1.196 &  0.100 &  $-$  &  $-$  &  n.c. \\
060 & 05:36:08.89 & -70:37:01.3 & 17.527 & 16.963 & 1.2757288 &  0.942 &   F  & 17.516 &  1.464 &  0.100 &  1.5  &  $-$  &   F   \\
061 & 05:36:24.32 & -71:30:39.4 & 18.393 & 17.745 & 0.8480731 &  0.632 &   F  & 18.389 &  1.049 &  0.110 &  1.2  &  $-$  &   F   \\
062 & 05:36:58.43 & -70:46:08.1 & 17.659 & 17.162 & 1.0590929 &  0.769 &   F  & 17.662 &  1.250 &  0.090 &  1.6  &  $-$  &   F   \\
063 & 05:37:54.39 & -69:19:28.7 & 18.756 & 18.013 & 0.8930306 &  0.534 &   F  & 18.762 &  0.849 &  0.180 &  1.0  &  $-$  &   F   \\
064 & 05:39:26.03 & -71:24:47.6 & 17.619 & 17.011 & 1.3574685 &  0.837 &   F  & 17.594 &  1.342 &  0.100 &  1.4  &  $-$  &   F   \\
065 & 05:40:03.04 & -70:04:47.8 & 17.508 & 17.041 & 1.3215432 &  0.888 &   F  & 17.515 &  1.386 &  0.130 &  1.3  &  $-$  &   F   \\
066 & 05:45:53.29 & -72:01:19.6 & 18.233 & 17.585 & 1.0400730 &  0.570 &   F  & 18.229 &  0.933 &  0.090 &  1.1  &  $-$  &   F   \\
067 & 05:48:22.07 & -70:45:49.3 & 18.451 & 17.786 & 0.8209206 &  0.629 &   F  & 18.407 &  1.082 &  0.100 &  1.2  &  $-$  &   F   \\
068 & 05:51:09.02 & -70:45:42.6 & 18.809 & 18.265 & 0.6256447 &  0.746 &   F  & 18.781 &  1.095 &  0.060 &  0.9  &  $-$  &   F   \\
069 & 05:53:31.91 & -71:32:44.8 & 17.460 & 16.823 & 1.5384361 &  0.674 &   F  & 17.454 &  1.134 &  0.080 &  1.5  &  $-$  &   F   \\
070 & 05:53:43.23 & -69:17:06.0 & 18.370 & 17.735 & 0.6293707 &  0.271 &  FO  & 18.354 &  0.471 &  0.100 & (2.2) &  1.3  &   F** \\
071 & 05:54:43.24 & -70:10:16.1 & 17.755 & 17.330 & 0.6762087 &  0.406 &  FO  & 17.755 &  0.687 &  0.070 &  $-$  &  1.2  &  FO   \\
072 & 05:59:08.23 & -68:24:01.2 & 17.535 & 17.038 & 1.0482162 &  0.901 &   F  & 17.530 &  1.385 &  0.030 &  1.6  &  $-$  &   F   \\
073 & 06:00:04.99 & -70:44:59.8 & 17.396 & 16.781 & 1.4650553 &  0.799 &   F  & 17.389 &  1.008 &  0.090 &  1.7  &  $-$  &   F   \\
074 & 06:02:45.18 & -70:45:31.2 & 17.551 & 16.889 & 1.5332247 &  0.322 &   F  & 17.549 &  0.498 &  0.070 &  1.7  &  $-$  &   F   \\
075 & 06:03:33.33 & -68:27:41.4 & 18.490 & 17.876 & 0.6920857 &  0.328 &   F  & 18.446 &  0.556 &  0.040 &  1.4  &  $-$  &   F   \\
076 & 06:04:33.02 & -70:57:55.8 & 17.671 & 17.032 & 1.5818249 &  0.563 &   F  & 17.669 &  0.874 &  0.070 &  1.2  &  $-$  &   F   \\
077 & 06:04:35.73 & -71:40:35.8 & 18.099 & 17.459 & 1.1224977 &  0.312 &   F  & 18.098 &  0.502 &  0.070 &  1.3  &  $-$  &   F   \\
078 & 06:06:58.20 & -72:52:08.7 & 17.417 & 16.979 & 0.8565557 &  0.391 &  FO  & 17.420 &  0.659 &  0.090 &  $-$  &  1.3  &  FO   \\
079 & 06:07:02.01 & -69:31:55.2 & 17.634 & 17.149 & 1.1551702 &  0.932 &   F  & 17.665 &  1.482 &  0.050 &  1.2  &  $-$  &   F   \\
080 & 06:09:35.12 & -70:10:42.1 & 17.997 & 17.336 & 1.0574863 &  0.307 &   F  & 17.995 &  0.540 &  0.090 &  1.6  &  $-$  &   F   \\
081 & 06:09:38.40 & -69:34:04.1 & 18.504 & 17.917 & 0.8008377 &  0.503 &   F  & 18.510 &  0.874 &  0.070 &  1.0  &  $-$  &   F   \\
082 & 06:16:58.31 & -70:52:18.9 &  $--$  & 17.207 & 0.7754317 &  0.469 & $-$  &   $-$  &   $-$  &   $-$  &  $-$  &  $-$  &  n.c. \\
083 & 06:18:52.45 & -70:51:55.4 & 18.236 & 17.646 & 0.5366171 &  0.497 &  FO  & 18.238 &  0.529 &  0.050 &  $-$  &  1.8  &  FO   \\
166 & 05:16:58.99 & -69:51:19.3 & 17.696 & 16.927 & 2.1105987 &  0.267 &  P2C  & 17.694 &  0.490 &  0.190 &  1.2  &  $-$  &  F   \\
\hline
\end{tabular}
\end{table*}
\normalsize


	\section{Comparison with theoretical predictions}\label{sec:comparison}

To properly compare theory and observations, we need to account for the
distance modulus and the reddening of the host galaxy. The distance modulus to 
the LMC has been extensively discussed in the literature, since this galaxy is the
cornerstone of the distance scale. We adopt here $\mu_0=18.50$~mag, as given
by a recent estimate using Red Clump stars observed in optical and near-IR 
\citep{fiorentino11}. This agrees with many of the measurements derived by independent 
methods, including classical Cepheids \citep{bono02}, RR Lyrae stars \citep{catelan08}, 
planetary nebulae luminosity function \citep{reid10}, the red clump, and the tip
of the red giant branch \citep{romaniello00}. The AC sample distributed
across the whole area surveyed by OGLE-III, we cannot adopt a unique 
reddening value for all the variables. To correct the magnitude of each
star for the appropriate extinction value, we used the reddening map 
recently derived by \citet{haschke11} using red clump and RR Lyrae stars
from the same OGLE-III release. 
In particular, we assumed the reddening value, $E(V-I)$, for the closest
position available for each individual AC. This is reported in column 11 of
Table~\ref{tab:tab1}. The mean reddening is $E(V-I) \sim 0.08$~mag, 
and only one star has a value higher than 0.2 mag.

\begin{figure}
\includegraphics[width=8cm]{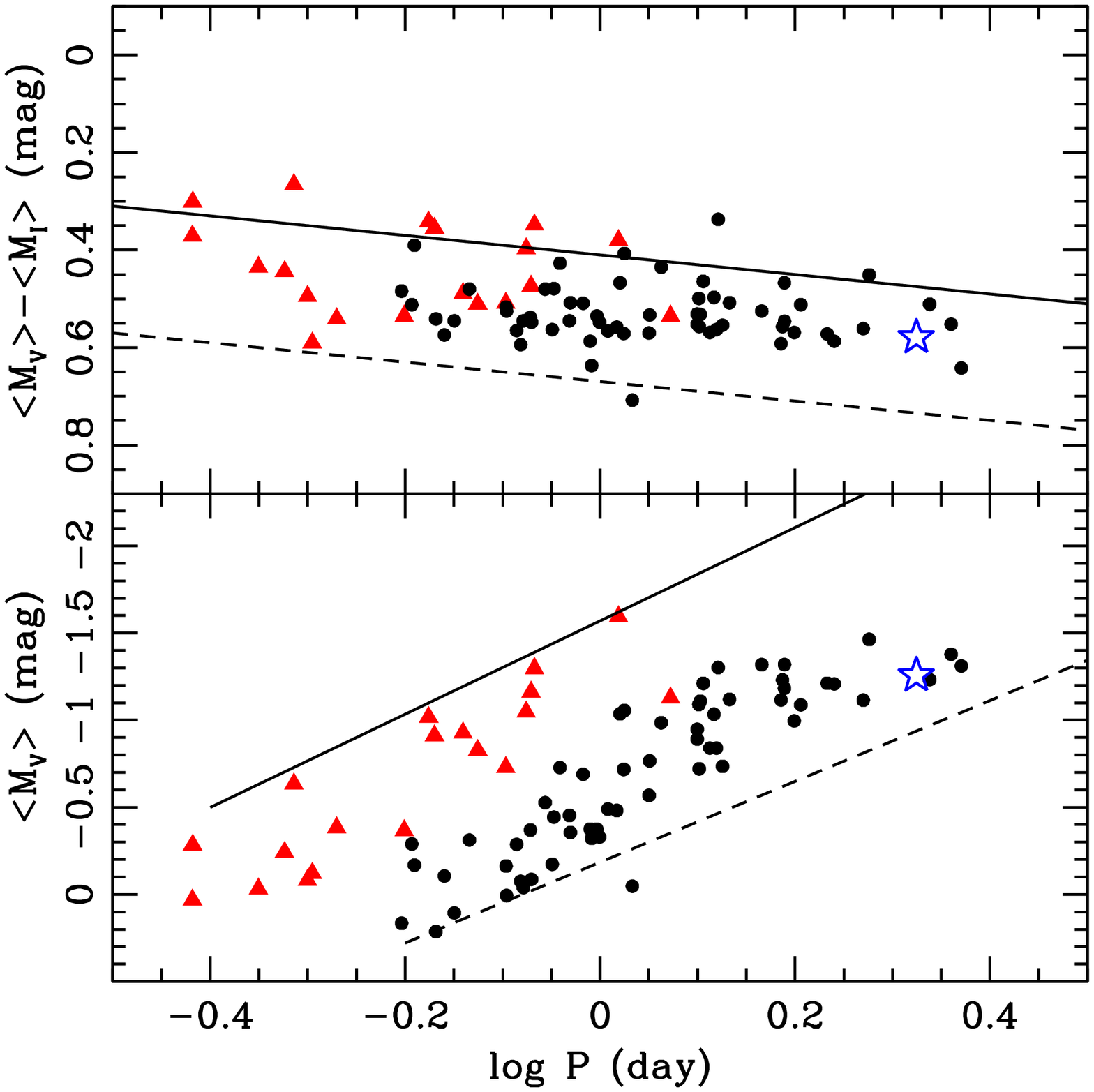}
\caption{Comparison between theoretical models and OGLE-III observations. 
The distance modulus assumed for LMC is $\mu_0=18.50$~mag
  \citep{fiorentino11}, and the individual reddening values are assumed
  using the reddening map given by \citet{haschke11} (see 
  text for details). The coding of the colours is the same as in previous
  figures. {\it Top --} ($<M_V>-<M_I>$)$_0$ colour vs period
  ($\log P$) diagram are compared with theoretical blue (solid line) and
  red (dashed line) edges of the instability strip \citep{marconi04}; 
  {\it Bottom --} Absolute magnitude ($M_V$) vs period ($\log P$) plane
  compared with the theoretical first overtone blue (solid line) and the
  fundamental red (dashed line) edges of the instability strip.
\label{fig:is}}
\end{figure}

In Fig.~\ref{fig:is} we show the comparison between the theoretical
prediction for the boundaries of the pulsation instability strip
\citep[][]{fiorentino06} and the sample of ACs. The top and bottom panels
show, as a function of the logarithm of the period, the absolute 
mean colour ($<M_V>-<M_I>$) and the absolute mean $<M_V>$ magnitude, respectively.
The agreement between theory and observations is good in both cases and 
for all the objects (the symbols are the same as in
Fig.~\ref{fig:wes}). The star, classified by OGLE-III as population
II Cepheids, seems to fully agree with AC observational
properties.
We note that there are few outliers
in the upper panel. This is possibly due to the uncertainty in the reddening
determination, with directly affects the true colour of stars. Nevertheless, this
comparison suggests that the correction for reddening applied is satisfactory for
most of the objects.

\begin{figure}
\includegraphics[width=8cm]{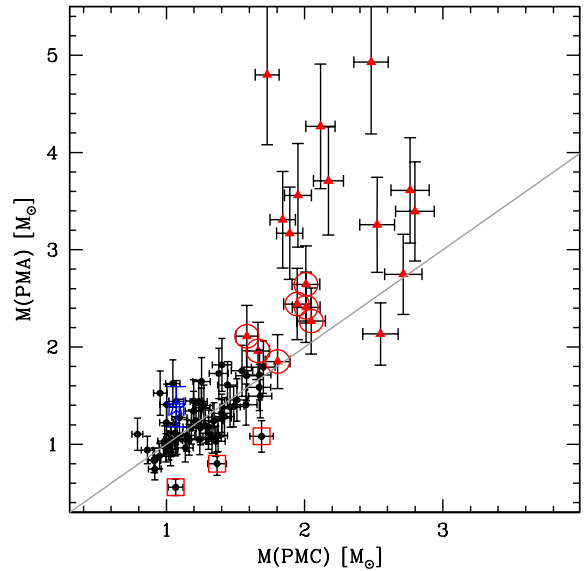}
\caption{Mass values for anomalous Cepheids in the LMC as derived from 
$PMA$ and $PMC$ relationships. Solid grey line shows the relation of
equality. The coding of the colours is the same as in previous
figures. The ACs identified as FO in OGLE but F in this work are
highlighted with red open circles, while the ones classified as F in
OGLE and FO in this work have been emphasized with open squares.
\label{fig:mass}}
\end{figure}

\begin{figure}
\includegraphics[width=8cm]{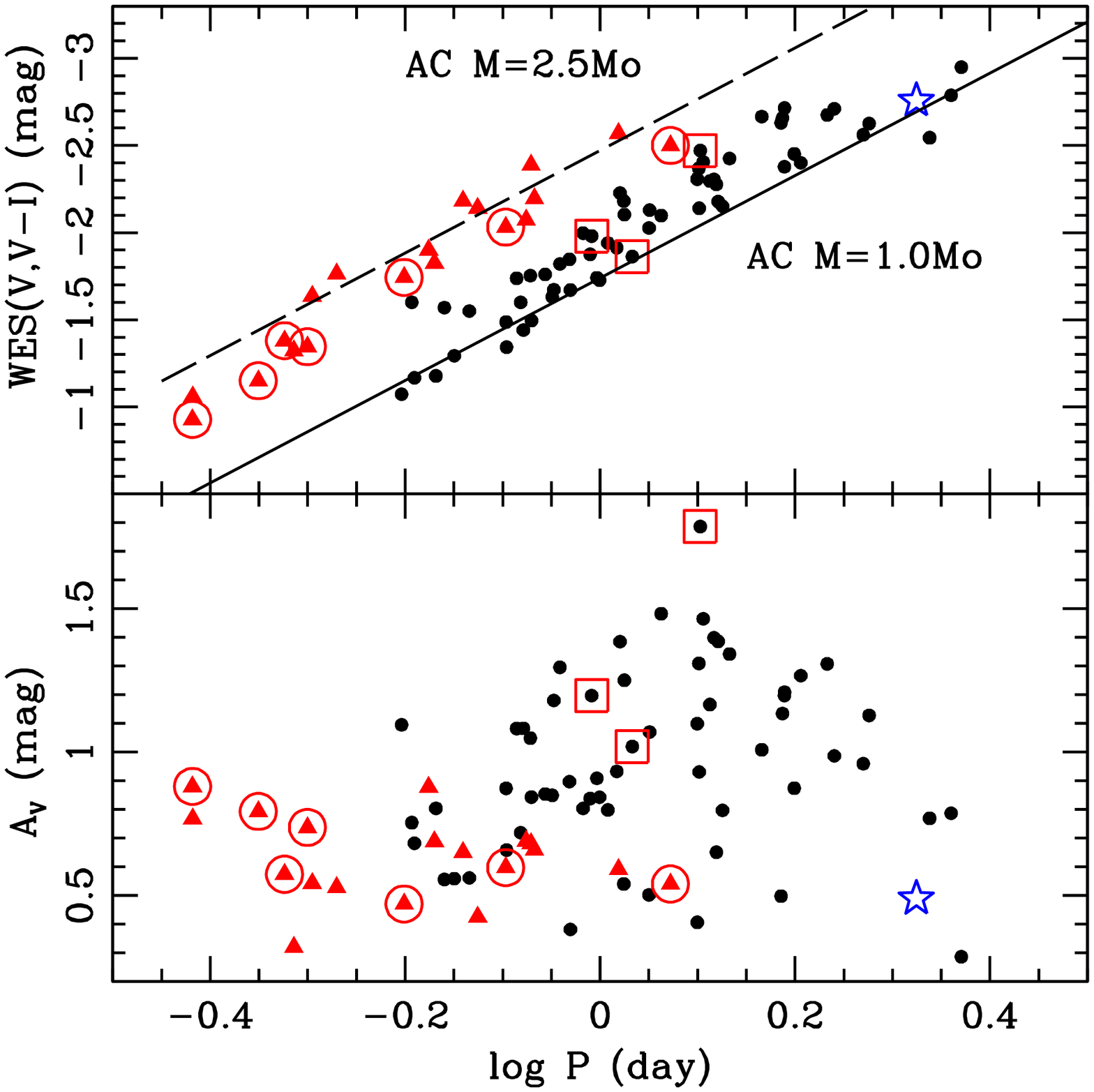}
\caption{{\it Top--} Same as Fig. \ref{fig:wes}, but we highlight the stars for
which we obtain a different mode classification as in Fig.~\ref{fig:mass}.
  {\it Bottom --} Amplitude ($A_V$) vs period ($\log P$) diagram. The coding of the colours
  is the same used in Fig.~\ref{fig:wes}. The amplitudes have been
  derived in this paper.
\label{fig:pav}}
\end{figure}


	\section{Constraining masses and pulsation modes}\label{sec:masses}

AC stars follow well-defined relations, which are the period-mass-amplitude relation 
($PMA$) and the mass dependent period-luminosity-colour relation (hereinafter $PMC$).
Here, it is worth 
mentioning that this theoretical scenario was used for classical Cepheids to 
constrain pulsation masses with very high precision (1\%) and returned values in 
good agreement with dynamical masses \citep[e.g.][]{pietrzynski10,pietrzynski11,cassisi11}.
In particular, it is possible to estimate individual masses once distance and 
reddening are known by applying the following equations, adapted from \citet{marconi04}:
\begin{equation}
M_{F,PMA}/M_{\odot} = 10^{(0.013-0.53~<M_V>- 1.3~\log P -0.244~A_V)} 
\end{equation}

\begin{equation}
M_{F,PMC}/M_{\odot} = 10^{(-0.97-0.53~<M_V>-1.55~\log P + 1.44 (<M_V>-<M_I>))} 
\end{equation}

\begin{equation}
M_{FO,PMC}/M_{\odot} = 10^{(-1.29-0.58~<M_V>-1.79~\log P + 1.54 (<M_V>-<M_I>))} 
\end{equation}
where $M_V$, $M_I$ are the absolute intensity-weighted mean $V$, $I$ magnitudes,
P is the period, and $A_V$ is the amplitude in the $V$-band. In this work, we 
assumed reddening law from \citet{cardelli89}. The formal uncertainty on the 
derived masses is 5\% using the $PMC$ and 15\% using the $PMA$ \citep{marconi04}. It is worth 
mentioning that the comparison of the masses derived with a different relation is 
independent of the distance modulus adopted, but it does depend on the reddening 
correction applied. Assuming the OGLE-III classification, we estimated the 
mass of each star, which is reported in Table~\ref{tab:tab1}. We 
provide the mean mass from Eqs. 1 and 2 in the case of F, and the mass from 
Eq. 3 for FO. For three objects, we were not able to estimate a reliable
mass, e.g. the mass estimated for the FU V-24 and V-59 with Eqs. 1 and
2 shows high discrepancies, and the mass estimated for FO V-28 is too high 
($\sim 2.3 M_{\odot}$) for an AC. 

A method is discussed in \citet{marconi04} which uses the previous relations
to simultaneously constrain the pulsation mode and the mass of the star. This 
method is based on the fact that the $PMA$ relation is only valid for the F pulsator, 
whereas the $PMC$ exists for both pulsation modes. Therefore, whenever the 
$PMA$ and the $PMC$ for F pulsators give consistent mass, this is a robust 
indication that the star is actually pulsating in the fundamental mode. The 
correct mode classification is important because it affects the mass determination. 
For the same luminosity and colour, we expect F pulsator to be more massive 
than FO pulsators. However, the usual diagnostics used for other classes of variable 
stars, such as the Bailey diagram ($A_V$ vs $\log P$ plane)
or the light curve morphology, do not allow the pulsation mode to be uniquely 
constrained. For this reason, we applied
the method given in \citet{marconi04} to check whether the mode classification based on the 
Wesenheit index (provided by the OGLE team) is consistent with theoretical 
pulsation predictions.

Figure~\ref{fig:mass} compares the mass values obtained using Eqs. 1 and 2. 
The grey line shows the relation of equality. Assuming the formal error for the
theoretical relations, we classified as F pulsators all the stars for which 
the mass difference is within the sum of the two $1-\sigma$ errors. For a better
identification, we have also classified F pulsators all those stars that did not 
pass the previous criterion, but show high $PMC$ masses ($\gsim 
2.5 M_{\odot}$), which are unrealistic in the adopted theoretical scenario. 
Out of 83 stars, we then classify 66 F and 17 FO, as reported in the last column 
of Table~\ref{tab:tab1}. Three stars (V-24, V-28, and V-59) present a
large mass difference and also an atypical amplitude ratio, and at least in one case,
a very red colour. This suggests that they might be blended sources, so they are 
reported as unclassified in Table~\ref{tab:tab1} (n.c.). Moreover, V-166 
previously identified as P2C turns out to be an AC pulsating in F mode.

The comparison with the OGLE-III classification reveals some discrepancy. Out of 19 FO 
for which we could estimate the pulsation mode, we only recover 12. The remaining seven 
stars were classified as fundamental mode pulsators with our method, and are labeled as 
``F**'' in Table~\ref{tab:tab1} and highlighted in Fig.~\ref{fig:mass}. For these 
stars, we also report, in parenthesis, the mass value corresponding to the F mode 
given by our classification. For these stars the FO mass value has a reliable value, 
systematically $\gsim 1 M_{\odot}$. Then, masses of these ``F**''  stars are 
compatible with both F pulsators with relatively higher mass, and FO pulsator of lower mass.

On the other hand, there are three stars, flagged with ``FO*'' in Table~\ref{tab:tab1} 
and classified  as FO, differently from the OGLE-III work. For these stars, the masses 
derived with the $PMC$ result to be always higher than the ones derived with the $PMA$ 
(less than $1 \sim M_{\odot}$), in disagreement with the general behaviour \citep{caputo04}. 
Their position in Fig.~\ref{fig:mass} shows that they are close to the bulk of F$-$mode stars.

To investigate the reason for these differences, we recalculated the masses of the 12
outliers varying the values of the reddening and the amplitude $A_V$ (which could
be affected by blending).
We found that, while modifying the amplitude has little or no effect, modifying the 
reddening value has a stronger effect. As an example, increasing the reddening of 0.1 
changes the classification of 2 FO (FO* in table) out of 3 to F and 5 F (F** in table) 
out of 7 to FO. This is an indication that the uncertainty on the reddening 
is the dominant source of error in the present analysis.

In Fig.~\ref{fig:pav} we show the Wesenheit relation, similar to Fig.~\ref{fig:wes}, 
(top) and the $P-A_{V}$ (bottom), and we highlight in both panels the stars for which 
we find a different mode classification, as in Fig.~\ref{fig:mass}. We note that 
the FO our method recovers as F all lie on the lower envelope of the FO 
relation in the Wesenheit plane, thus supporting the classification's consistency
with both FO and F modes. Also, the Bailey diagram clearly shows that, as opposed 
to the case of RR Lyrae stars, this plane is not an efficient diagnostic to separate F and 
FO pulsators. In fact, while the shortest period ACs are clearly FO, while those 
with large amplitudes tend to be F, there is a clear degeneracy in the low-amplitude
regime, for periods with  -0.2$\lsim \log P \lsim$ 0.1~d.

\begin{figure*}
\includegraphics[width=17cm,height=9cm]{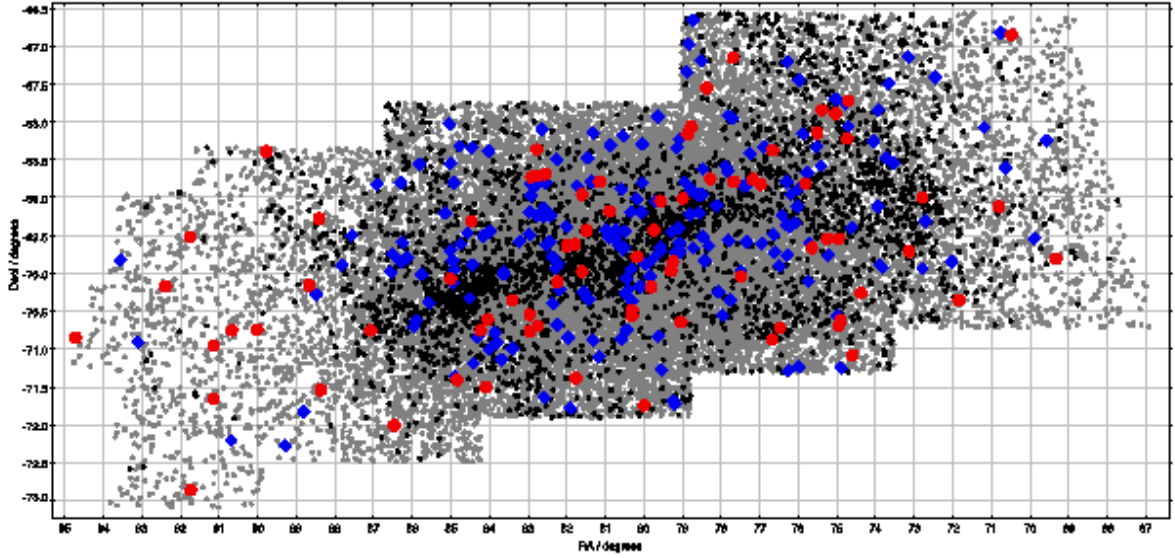}
\caption{Spatial distribution of  anomalous (red), population II
  (blue), and classical Cepheids (black), RR Lyrae (grey) in the total
  area  of the Large Magellanic
  Cloud ($\sim 39.7$~square degrees) covered by OGLE III survey. 
\label{fig:radec}}
\end{figure*}

\begin{figure}
\includegraphics[width=8cm]{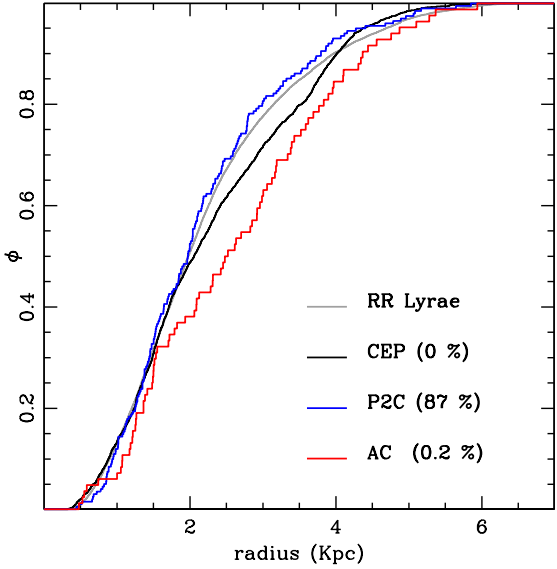}
\caption{Cumulative distributions for anomalous (red), population II
  (blue), and classical Cepheids (black), RR Lyrae (grey).
\label{fig:cumul}}
\end{figure}


	\section{Spatial distributions of variable stars}\label{sec:spatial}
	
To better understand the origin of ACs we take advantage of the extensive 
and complete sample of variable stars discovered by OGLE-III survey of the
LMC, namely 22,651 RR Lyrae ($ab$ and $c$ type), 203 P2C, and 3,087 young 
classical Cepheids (F and FO only). The Magellanic clouds are the only 
galaxies where such an extensive variable star study has been performed, 
covering variables with a wide range of periods. The total area covered 
of the LMC by the OGLE-III project is 39.7 square degrees and samples the 
central regions of the galaxy. In particular, it covers the bar and the inner 
disc out to a maximum distance of $\sim 5$~kpc. However, the distribution of 
OGLE-III fields follows the bar direction from north-west to south-east.
Also we note that the size of the LMC is significantly greater than the area 
covered by OGLE-III \citep[e.g. ][]{saha10}. Nevertheless, the present data 
set allows studying the spatial distribution of the different samples of variable 
stars over the central region of the galaxy, thus tracing the parent population. 
As a matter of fact, the RR Lyrae stars and the P2Cs are horizontal branch stars 
tracing the old and predominantly metal-poor population ($\mathrm{[Fe/H]} \lsim -1$,
~dex), while the classical Cepheids trace the young and mainly metal-rich 
($\mathrm{[Fe/H]} \gsim -0.5$~dex) stars. As shown in Fig.~\ref{fig:radec}, the spatial
distribution of young stars is concentrated at the centre of the
galaxy, nicely tracing the bar region where the star formation occurred
at very recent epochs ($\lsim 0.5$\,Gyr ago). Also, there are two clear
overdensities at the end of the bar at $\sim 85$ and $74$~degrees. On the other
hand, the RR Lyrae stars seem to follow a much broader uniform distribution as
is usually observed in other galaxies \citep{monelli10a,monelli10b,monelli12a,deboer11}. 
Interestingly, the ACs do not seem to obviously follow either of the two previous
distributions. In particular, while the majority seems concentrated in the innermost
regions, where the bulk of the RR Lyrae stars is located, they are not usually
placed in the bar. Moreover, there is a clear overabundance in the east direction
($RA \gsim 87$~degrees), where the density of both RR Lyrae and Cepheids has 
significantly decreased. A similar excess does not seem to be present on the other 
side of the galaxy. 

\begin{figure}
\includegraphics[width=8cm]{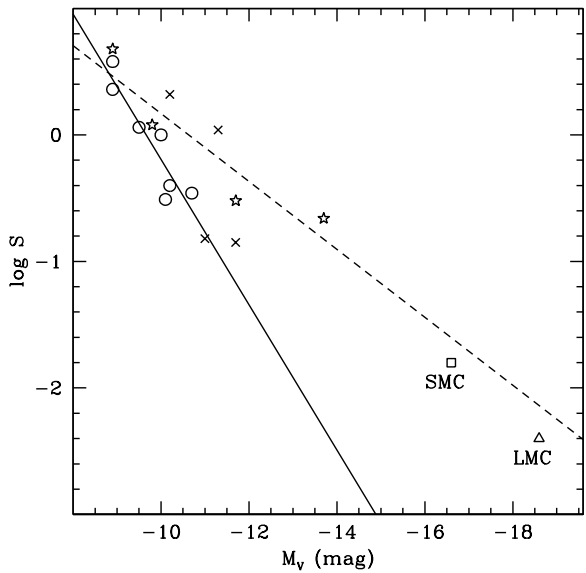}
\caption{AC frequency ($\log S$) per $10^5 L_{\odot}$ as a function of the absolute 
$V$ magnitude of the host galaxy ($M_V$). Open circles show purely old dSph galaxies, while starred symbols indicate galaxies
with large intermediate-age populations. The crosses mark the four M31 satellites,
The LMC and SMC are marked by the open triangle (for an assumed luminosity of 
$2 \times 10^9 L_{\odot}$) and square, respectively. The  
solid line show the fit to the relation for the old systems only, while the dashed 
line is the fit to the four crosses. We note that the present value for the LMC is a lower limit, 
since ACs have so far been detected only in the central region of the LMC. However, 
the LMC is known to be significantly larger, possibly occupying an area $\sim 9$ times 
larger \citep[see e.g.][]{irwin91, majewski09, saha10}.
\label{fig:frequency}}
\end{figure}

To better compare the spatial distribution of the different class of variable stars,
we built the cumulative distributions as a function of the galactocentric radius 
using the LMC model from \citet{vandermarel01} to derive the deprojected distance 
from the centre. They are shown in Fig.~\ref{fig:cumul} for the RR Lyrae stars, classical, 
P2Cs, and ACs. We have also computed, using a Kolmogorov-Smirnov test, 
the probabilities that the various sample are drawn from the same parent population
as the RR Lyrae stars. We found that this is 0\% ($5\times 10^{-9}$\%) for classical Cepheids,
0.2\% for ACs, and 87\% for P2C. Therefore, the P2Cs seem to follow the same distribution 
of the RR Lyrae stars, in agreement with the fact that they both originate in the old
population. We have also computed the probability that the ACs come from young classical
Cepheids, which is 8\%. To avoid possible bias from the area covered by the OGLE-III
survey, we repeated these estimations using only subgroups of stars located at distances
within 2, 3, and 4~kpc from the LMC centre. We find that the probability that the 
distributions of the RR Lyrae and P2C coincide is always above 75\%, while that of RR Lyrae
and classical Cepheids is always below 1\%. In the case of ACs and RR Lyrae stars, 
the probability widely fluctuates but, with some hint of decreasing probability
for increasing the limit radius, ranging from 32, to 10, to 0.5\% in the three cases.

These tests do not show any clear similarity between the distributions of ACs with 
either that of classical Cepheids or RR Lyrae stars. This suggests that they are not
uniquely correlated either with the old or with the youngest LMC population.

	\section{Discussion of the origin of ACs}\label{sec:discussion}

The theoretical scenario that explains the structure, evolution, and pulsation 
properties is settled for different classes of pulsating variable stars. 
This has strong implications in the study of the star formation history of 
galaxies, because it is possible to characterize the properties of the parent 
population (e.g., age and metallicity) by investigating the bright variables associated. 
However, the presence of ACs does not uniquely identify the nature of their
progenitors. The two most favoured formation channels are the evolution of single, 
metal-poor star with mass $\lsim 2M_{\odot}$, or the evolution of coalescent binary
systems of metal-poor stars. Independent of the formation mechanism, a crucial 
characteristic of the progenitor for producing an AC appears to be its low 
metallicity. In fact, central-helium burning structures in the AC mass range, but 
more metal-rich than $Z \sim 0.0008$ \citep{gallart04b,fiorentino06}, are too 
red to enter the Cepheid instability strip and populate the red clump.
To have pulsators at higher metallicity, the mass of the progenitor must be higher.
In this case, the star will cross the strip at higher luminosity
producing short period 
classical Cepheids, which ignite the helium quiescently. In particular, the minimum mass 
of central helium-burning pulsators crossing the instability strip increases from 
$1.9 M_{\odot}$ (for $Z=0.0004$) to $3.6 M_{\odot}$ (for $Z=0.008$, \citealt{caputo04}).

The binary origin is a plausible explanation of ACs having been ubiquitously 
discovered in low-density, very metal-poor environments ($Z \le 0.0008$), such as
nearby purely old dwarf galaxies. In fact, ACs have been observed in dwarf 
galaxies that do not show any intermediate-age population, such as Draco, Ursa Minor, 
Sculptor, Sextans, Tucana, LeoII, and Cetus. This implies
that, at least in these galaxies, ACs must be the progeny of 
metal-poor, binary systems. On the other hand, very few systems have been discovered to
host both classical and anomalous Cepheids. \citet{gallart04} first pointed out
this occurrence in the case of the transition dIrr/dSph galaxy Phoenix, supporting
the evidence that the same applies to Sextans~A and Leo~A. Comparing the spatial 
distribution of different samples of stars in Phoenix, \citet{gallart04} find that,
while classical Cepheids are strongly segregated in the central regions, following 
the spatial distribution of the youngest population, the ACs are distributed over a 
broader area, but are still more concentrated than the RR Lyrae stars, which can be safely 
taken as representative of the old population. This suggests that the ACs in Phoenix
are mostly intermediate-age stars, though the possibility that a few are BSS descendents
cannot be excluded. This implies that, despite its prolonged star formation until epochs 
that are recent enough to produce classical Cepheids, Phoenix experienced little chemical 
enrichment, in agreement with \citet{hidalgo09}, and the $1-6$\,Gyr old population was 
still metal-poor enough to produce ACs.

		\subsection{The case of LMC}\label{sec:origin}

The case of the LMC presented here is probably more complex, since the LMC is two orders
of magnitude more massive than the typical satellite dSph, and has a complex structure 
(i.e. disc and bar). Nevertheless, until the work by \citet{soszynski08c}, no ACs had 
been unambiguously detected. The present sample of 84 objects is the largest so far 
for any external galaxy. The coexistence with large samples of different variable 
stars allowed us to compare the radial distribution of these objects using a
Kolmogorov-Smirnov test. We found that the present sample of ACs cannot be associated 
to the young classical Cepheids. It must be stressed here that given the large sample 
of AC and classical Cepheids in the LMC, their different nature clearly appears for 
the first time. Despite ACs occupy the same instability strip of the classical Cepheids 
as suggested by \citet{caputo04} and populate the low-mass and low-metallicity extension,
it is clear that the two groups follow different Wesenheit relations \citep{soszynski08c}.
We also found that the radial distribution of ACs do not correlate with the oldest 
population, as represented by RR Lyrae and P2C, suggesting that, as for of case of 
Phoenix \citep{gallart04}, a large fraction of single intermediate-age stars is present.

To investigate this hypothesis, we interpreted the properties of ACs in the full
picture of the LMC star formation history. Given the mass derived in this work 
(see Sect.~\ref{sec:masses}), which is $1.2\pm 0.2 M_{\odot}$ for the bulk of ACs,
we can estimate the corresponding age. Using scaled solar evolutionary
tracks from the BaSTI\footnote{http://albione.oa-teramo.inaf.it/}
\citep{pietrinferni04} 
database, we find ages from 1.8 to 5.8\,Gyr, assuming Z=0.0006. However, the age-metallicity
relation (AMR), based on spectroscopic analysis, as presented in 
recent years \citep{cole05,carrera08a,carrera11}, reveals that the mean metallicity in this 
age range is an order of magnitude greater than the expected maximum value for ACs, 
being close to Z=0.008. Interestingly, the spatial analysis by \citet{carrera08a}
and \citet{carrera11} reveals that, in a number of disc fields at different distances
from the LMC centre, but also in the bar field from \citet{cole05}, the metallicity
distribution always presents a well-populated tail toward low
metallicity ($Z\sim 0.0004$),
but the most metal-poor stars are systematically older than 10\,Gyr. 
This occurrence seems at odds with our hypothesis that the bulk of ACs are metal-poor
intermediate-age stars. However, one could ask whether, due to the limited number of 
spectroscopic targets so far analysed, we are facing an observational bias, and that
very few metal-poor stars exist in the age range from $\sim 2-6$\,Gyr.   

We can attempt to estimate the percentage of the mass involved in the
star formation episode that generated the ACs, following the simple prescriptions
given in \citet{renzini88}, which hold for each post-main sequence
phase. The number of ACs ($N_{AC}=84$) observed is related to the
total luminosity ($L_{Z<0.0008}$) of the population they belong to via the following 
relation: $N_{AC}=B(t) L_{Z<0.0008}\tau$. Where B(t) is the specific evolutionary 
flux \citep[$\sim 0.15 \times 10^{-10}$ stars per $L_{\odot} yr^{-1}$, see][]{renzini88},
$\tau$ is the time spent in the instability strip for a mass of about $1.2 
M_{\odot}$ \citep[$\sim 50$~Myr, see][]{fiorentino06}, and $L_{Z<0.0008}$ results 
to be $1\times 10^5 L_{\odot}$. This value is very low related with the total luminosity 
of the LMC ($L_{TOT} = 2 \times 10^9 L_{\odot}$), even if we account for the limited area 
covered by the OGLE-III survey. Therefore, the corresponding expected total number of 
stars from this population is very low, a few thousand over the whole field-of-view, 
implying that few stars brighter than the Horizonal Branch are expected per square degree.
This strongly suggests that such a small event cannot be properly sampled by the SFH 
analysis or spectroscopic investigation of incomplete samples. This in turn highlights
the key role of ACs in this context. In fact, the discovery that a small amount of gas
was still poorly enriched 6\,Gyr ago gives additional 
constraints to the star formation history that have to be accounted for in the formation 
and evolution modeling of the LMC. 

Finally, we want to briefly comment the possibility that ACs are young, but 
metal-rich stars looks unfeasible. Stars with metallicity close to $Z=0.008$ are 
expected to be significantly more massive to enter the instability strip during the 
central helium-burning phase, with respect of more metal-poor stars. 
Even if we invoke unusually high mass loss to reconcile with the low-mass we
estimated in this work, the mass of the helium core would not
change significantly, leaving us with brighter stars, which would be at odds with
the observed properties of ACs.

		\subsection{Comparison with other galaxies}\label{sec:compa}

ACs appear to be intrinsically rare objects. This is valid in general for nearby dwarf 
galaxies and for the LMC in particular. The OGLE-III projects surveyed more than 35 million 
stars, and only 84 ACs were detected. In this case, we can safely assume that the 
completeness is very high, both because of the large number of measurements and the 
photometric quality of the data. 

Nevertheless, \citet{mateo95} first notice that a correlation exists between the frequency 
of ACs ($\log S$, per $10^5 L_{\odot}$) and the total luminosity of the
host galaxy ($M_V$), valid 
for many nearby satellites. In particular, it was found that the frequency of ACs 
decreases for increasing luminosity of the host galaxy. \citet{pritzl02} noticed that the 
surveyed M31 satellites also
follow this trend, as does Phoenix \citep{gallart04}. Figure~\ref{fig:frequency} shows
the frequency of ACs as a function of the absolute visual magnitude of the host galaxies.  We 
used the data from Table 5 of \citet{pritzl04}, to which we added the values for Phoenix 
\citep{gallart04}, Cetus (\citealt{bernard09}, \citealt{monelli12a}), Tucana 
\citep{bernard09}, Andromeda I, and Andromeda III \citep{pritzl05}. The symbols used in the 
figure differentiate the objects according to their SFH (see the caption for details). In particular,
we represent purely old systems (i.e. Ursa Minor, Draco, Sculptor, Leo~I, Sextans, 
Tucana, and Cetus) and dwarf galaxies characterized by an important intermediate-age population 
(Carina, Fornax, Phoenix, and LeoII). We also plot four M31 satellites (And I,
And II, And III, and And VI), for which the data available are 
not deep enough to set tight constraints on their SFH. In fact, the photometry reaches 
roughly one mag below the horizontal branch, and it is impossible to clearly detect the
presence of a strong intermediate-age population 
\citep[for example, compare with Carina in Fig. 1 of][]{harbeck01}. 

The two lines are the fits to the two group of galaxies. The plot suggests that purely 
old systems and intermediate-age population 
systems follow different relations. In particular, when an intermediate-age population is present,
the frequency of ACs tends to be higher than expected from a purely old population, and the 
discrepancy increases for increasing luminosity. The underlying physical mechanism of such a
relation is unclear. However we only expect ACs resulting from the evolution of binary systems
in purely old systems. Thus, the correlation with the luminosity, hence the mass,
of the host galaxy may indicate that binary systems have higher chances of surviving
in low-mass systems. It is worth recalling in this context that a similar correlation has 
been found for BSSs. In fact, \citet{momany07} have first shown that the frequency of BSS with
respect of the number of horizontal branch stars decreases for increasing luminosity of the
host galaxy. This might be an independent evidence that an evolutionary link exists between
the binary systems, BSSs, and ACs. If such a correlation could be explained this way for 
purely old systems, the fact that galaxies known to have intermediate-age population
present an excess of ACs, compared to purely old galaxies, would 
be simply due to the coexistence of ACs formed via both formation channels. Among 
the four M31 satellites, two follow the fit of the old dSphs (And II, And III), while two 
present an apparent excess of ACs (And I, And VI). This is possibly
because the small fraction of these galaxies covered by the data for these
galaxies introduces some bias, but it is also possible that And I and And VI host some 
intermediate-age population, at present undetected but responsible for this effect. 

What seems really surprising is that both Magellanic clouds appear to follow the same
relation of these four galaxies, but in a significantly higher luminosity regime.
It is also interesting to note that the SMC hosts very few ACs candidates 
\citep[3 F and 3 FO][]{soszynski10a}. As for to the LMC, it presents an excess of ACs, 
thus supporting that it is possible that a small population of intermediate but metal-poor
stars exists.

The new OGLE-IV survey will possibly shed light on this problem. In fact, it is well known
that a significant age gradient is present in the disc of the LMC \citep{gallart08,carrera11} 
in the sense that the age of the youngest population gets older for increasing radius. This 
means that, starting from a certain distance, we do not expect any population that is young
enough to produce ACs, thus leaving only the BSS aftermath, if present. Moreover, the absence
of young main sequence stars will permit direct detection of BSSs, thus allowing a direct
correlation with ACs. The distance where this is expected to occur is far beyond the 
present coverage of the OGLE-III survey, but it will be definitely covered by OGLE-IV,
at present in execution. Stronger constraints on the formation mechanisms of the ACs in the
LMC are therefore expected in the near future. Also, a systematic investigation of the M31
satellites to derive the details of their SFH, which is at present perfectly feasible
for the ACS and WFC3 cameras on-board the Hubble Space Telescope, would help  solve this 
problem.


	\section{Conclusions}\label{sec:conclusions}

We have analysed the pulsation and spatial properties of 83 ACs discovered by the OGLE-III survey 
in the LMC. The main results obtained are the following:

$\bullet$ We classified one star (V-166), previously identified as P2Cs, as bona-fide ACs, thus 
increasing this total number to 84 stars.

$\bullet$ We performed the Fourier analysis of the $V$-band data, deriving the $V$-band amplitude
$A_V$ and mean $V$ values for the 83 stars for which the $V$
photometry is available, with only one excluded (V-82). We then 
applied the method introduced by \citet{marconi04} to simultaneously derive individual masses and
pulsation modes. We find good agreement with the OGLE-III
classification for most of the stars ($\sim 85$\%), based on the Wesenheit relation
and the morphology of light curves. The mass of the ACs ranges from 0.8 to 1.8 $M_{\odot}$, 
with a mean value of $1.2 \pm 0.2 M_{\odot}$.

$\bullet$ The large sample of ACs allowed, for the first time, to unambiguously show that anomalous and
classical Cepheids are different. In fact, even though they populate the same instability strip,
they follow distinct period-luminosity relations and different spatial distributions.

$\bullet$ We compared the spatial distribution of the ACs with that of other samples of variables
(RR Lyrae, P2C, classical Cepheids) using a Kolmogorov-Smirnov
statistical test. We found that RR Lyrae and P2C, both bona-fide representatives
of the old population ($>10$\,Gyr) follow the same distribution, which is different from that
of the young classical Cepheids. The distribution of ACs does not resemble any of the other samples.
We conclude that a large fraction of the ACs descend from single, metal-poor,
intermediate-age stars that are $1-6$\,Gyr old.

$\bullet$ The comparison with the presently available SFH and AMR shows that such a population
is negligible. We show that the total mass formed in that star
formation event is about $5\times 10^{-5}$ the total mass of the LMC, thus resulting in very few stars that are very unlikely
to be observed in spectroscopic surveys or in statistical approaches to derive the SFH.
This shows the importance of variable stars and rare objects like ACs for giving independent
constraints on the modeling of galaxy evolution.

$\bullet$ We reanalysed the frequency-luminosity relation discovered by \citet{mateo95}, showing
that, if only purely old systems are considered, the correlation is tighter and steeper. The objects
that present important intermediate-age population tend to larger AC frequency, suggesting
that ACs formed from single stars are also present.

The detection and characterization of ACs at much greater distances from the LMC centre
would be very interesting. In fact, given the strong gradients in the LMC stellar
population, starting from $\approx 8$~Kpc the young population completely disappears. The detection
of ACs in these regions of the LMC would strongly point to the binary origin.


\section*{Acknowledgments}
 
We sincerely thank the anonymous referee for his/her pertinent comments
that improved the readability of this paper. We warmly thank Igor Soszy{\'n}ski 
for sharing the individual $V$-band photometry; Eline Tolstoy, Giuseppe Greco, 
Marcella Marconi, Michele Bellazzini, and Ricardo Carrera are warmly thanked for 
useful discussions and support. Carme Gallart, Edouard Bernard, Giuseppe Bono, and 
Santi Cassisi are appreciated for reading an early draft of this manuscript. 
GF has been supported by the INAF fellowship 2009 grant.
MM acknowledges the support by the IAC (grant 310394), the Education and
Science Ministry of Spain (grants AYA2007-3E3506, and AYA2010-16717)

\bibliographystyle{aa} 
\bibliography{my}

\end{document}